# Benchtop Fabrication of Memristive Atomic Switch Networks

Henry O. Sillin[1†], Eric J. Sandouk[1†], Audrius V. Avizienis[1], Masakazu Aono[2], and Adam Z. Stieg[2,3,*] and James K. Gimzewski[1,2,3]

[1]Department of Chemistry and Biochemistry, University of California, Los Angeles, CA 90095, USA
[2]WPI Center for Materials Nanoarchitectonics (MANA), National Institute for Materials Science (NIMS), Tsukuba, Ibaraki 305-0044, Japan
[3]California NanoSystems Institute, University of California, Los Angeles, CA 90095, USA

**Abstract.** Recent advances in nanoscale science and technology provide possibilities to directly self-assemble and integrate functional circuit elements within the wiring scheme of devices with potentially unique architectures. Electroionic resistive switching circuits comprising highly interconnected fractal electrodes and metal-insulator-metal interfaces, known as atomic switch networks, have been fabricated using simple benchtop techniques including solution-phase electroless deposition. These devices are shown to activate through a bias-induced forming step that produces the frequency dependent, nonlinear hysteretic switching expected for gapless-type atomic switches and memristors. By eliminating the need for complex lithographic methods, such an approach toward device fabrication provides a more accessible platform for the study of ionic resistive switches and memristive systems.

**Introduction.** As ongoing trends in device technology continue on a path toward operation at increasingly reduced spatiotemporal and energetic scales, the perpetual demand for increased density in solid-state electronics and integrated circuits requires new approaches in device fabrication. Modern approaches to advanced computation commonly involve solid-state very-large-scale integration (VLSI) circuits which increase the density of functional elements at reduced dimensionality through a merger of complementary metal–oxide–semiconductor (CMOS) technologies with nanoscale architectures known as CMOS-Molecular (CMOL) [1]. As a result, the fabrication of nanoelectronic devices has typically been consigned to advanced techniques and sophisticated equipment.

Realization of the nanoscale memristor [2, 3], an exciting class of two-terminal circuit elements whose behavior can be defined as a relationship between charge and flux [3-5], has shown great promise in circumventing the challenges of CMOS-based electronics. The memristor is in theory as fundamental to electronic circuit design as are the resistor, capacitor, and inductor [6]. Functional memristive devices show both volatile and non-volatile memory capacity, while current fabrication methods allow for nanoscale memristive elements to be integrated in ultra-high dense topographies [7-9].

In practice, nanoscale memristive devices generally exist as a metal-insulator-metal (MIM) junction consisting of two conductive electrodes separated by an insulating gap of width, $D$. Application of an external bias voltage alters the conductance of the insulator by various mechanisms including but not limited to charge carrier migration, phase changes and magnetic domain rearrangements [2, 10-13]. In the case of a charge carrier based device, devices operate in an ionic drift model whose state equation modifies the constant parameter of resistance, $R$, in Ohm's law $V(t)=RI(t)$, with the charge ($q$) dependent memristance, $M(q)$:

$$M(q) = \left[\frac{\mu_v R_{ON}^2}{D^2} q(t) + R_{OFF}\left(1 - \frac{\mu_v R_{ON}}{D^2} q(t)\right)\right] \quad (1)$$

where $\mu_v$ is the charge carrier mobility, $R_{ON}$ and $R_{OFF}$ are the resistance states for the maximal and minimally doped states, respectively, and $D$ is the gap width [2]. Through variation of the gap width, memristive systems act similarly to leaky transistors that can be controllably set to different resistance states in a continuum between high and low resistance states ($R_{OFF}$ and $R_{ON}$, respectively) as a function of applied voltage.

Atomic switches are a class of nanoscale electroionic circuit elements that exhibits memristive switching under applied AC bias through formation and annihilation of a metallic filament as seen in Figure 1 [10, 14, 15]. In the specific case of a silver sulfide atomic switch, the MIM junction is composed of Pt, W, or Ag conductors, and a $Ag_2S$ insulator which serves as an ionic conductor. Silver sulfide and has two distinct phases – insulating α-phase acanthite ($2.5 \times 10^{-3}$ $\Omega cm^{-1}$) at room temperature and conductive β-phase argentite ($1.6 \times 10^{3}$ $\Omega cm^{-1}$) above 178° C. It is known that external bias voltage causes the α-$Ag_2S$ insulator to transition into the conducting high temperature β-phase [16]. Silver cations are more mobile in argentite, so this phase transition facilitates a migration of $Ag^+$ ions towards the cathode where they are reduced, forming a highly conductive Ag filament [17-19]. When negative bias is applied, the Ag filament oxidizes, β-$Ag_2S$ reverts to α-$Ag_2S$ and $Ag^+$ ions re-dissolve into the insulator. Thus the conducting channel is broken and the junction returns to the high resistance state. Under repeated bipolar voltage, the filaments continually reform and re-dissolve, resulting in repeatable resistive switching.

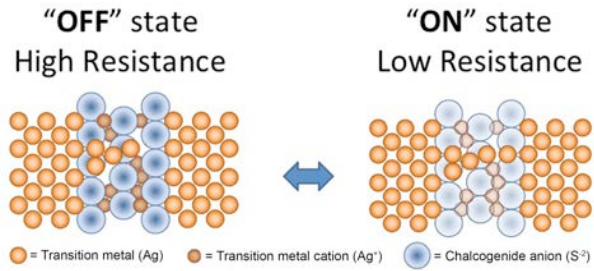

**Figure 1 - Operational schematic of a Ag|$Ag_2S$|Ag atomic switch**. Changes in device resistance result from the formation of a conductive Ag filament across the insulator. In the OFF state the Ag channel is incomplete as it is separated by the insulating $Ag_2S$ layer. Applied voltage stimulates the migration of $Ag^+$ cations to the end of the filament, extending it and completing the channel, which characterizes the ON state.

Due to their small size, low power consumption, and nonlinear characteristics memristive circuit elements are rapidly emerging as a complementary technology to CMOS-based computation and memory storage. However, the development of commercial devices based on memristive elements remains limited due to the factor of $1/D^2$ in eqn. (1), which causes the nonlinear memristive behavior to be dramatically stronger at small values of $D$. Devices which display nonlinear resistance at the voltage, current, and time scales of conventional CMOS transistors typically require an insulator gap size at the nanometer scale. As a result, precision nanofabrication techniques [2, 7], scanning probe microscopes [10], pulsed laser [20] and atomic layer deposition [21] are employed to achieve such gap widths.

In contrast to complex lithographic methods, solution-phase approaches have been shown to produce functional memristive devices using spin-coated solgel films alongside anodic electrochemical deposition [22-29]. Electrochemistry in particular offers a divergent approach for the fabrication of metallic structures. Specifically, electroless deposition of various metals through the spontaneous reduction of soluble metal cations is a mature technology that has been employed extensively in macroscopic plating applications and the manufacture of printed circuit boards. Under diffusion-limited conditions [30], the reaction has been shown to generate a diversity of self-assembled structures including nanowires, dendrites, and fractals at the micro- and nanometer scales [31-34]. Such assemblies present an opportunity to readily achieve MIM interfaces with sufficiently small gap sizes to enable memristive operation.

Here we present the fabrication of functional memristive devices based on a self-assembled network of highly-interconnected Ag|$Ag_2S$|Ag atomic switches using entirely inexpensive materials and techniques following the scheme shown in Figure 2. These atomic switch networks (ASN) rely on the electroless deposition of silver from copper seeds to form an intricate fractal wiring architecture suitable for creating a network of nanoscale MIM junctions. Applied bias converts these junctions into atomic switches and the entire network demonstrates the frequency dependent, nonlinear hysteretic switching requisite for applications in data storage and computation. Our approach drastically reduces the need for precision fabrication or placement of the switching junctions as compared to typical lithographic methods. This demonstrates the utility of alternative fabrication techniques, and provides a unique approach toward the production of functional memristive devices. The process described here has further implications for the fabrication of large memristive devices and networks, opens future development to a wider field of investigation, and facilitates the study of memristive devices in educational environments.

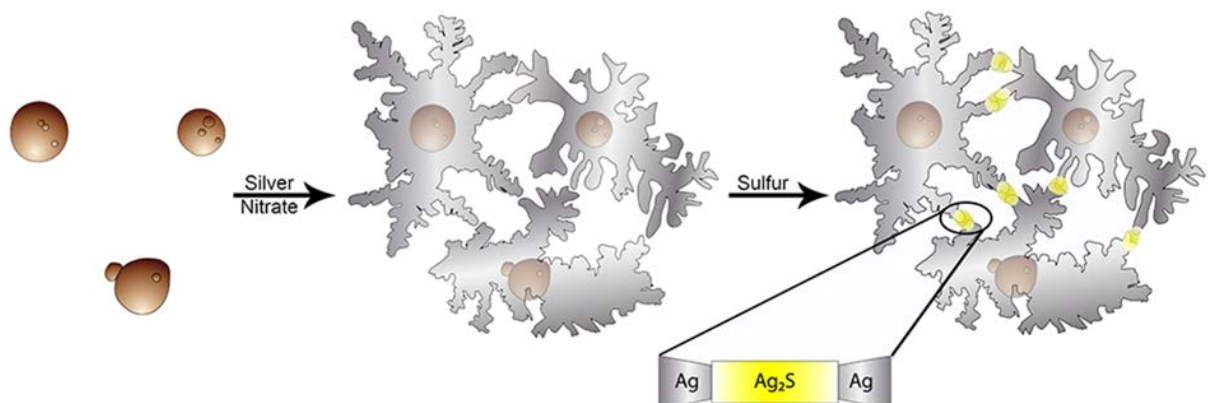

**Figure 2 - Schematic for fabrication of fractal ASN by electroless deposition.** Copper microsphere seeds serve as nucleation sites for the spontaneous reduction of soluble silver ions and formation of dendritic silver structures which are subsequently converted to $Ag_2S$ through gas phase sulfurization. The $Ag_2S$ layer creates the MIM interfaces required for resistive switching.

**Experimental Methods.** Atomic switch network devices were fabricated using inexpensive and commonly available materials, and followed by self assembly processes to create an interconnected network of nanoscale MIM junctions. Two silver coated copper wires were fixed to a glass microscope slide with kapton tape with a 2-4 mm separation between wires (Figure 3a). Next, a solution of copper microspheres (d = 10 µm, 99.995% purity, Alfa-Aesar) was prepared in isopropyl alcohol (5 mg/mL). A small droplet (20 µL) of this solution was transferred via pipette between the wires, allowing the Cu spheres to disperse uniformly across the region (Figure 3b). Following solvent evaporation, a droplet of 50 mM aqueous $AgNO_3$ was dropcast onto the Cu treated areas. With the Cu spheres serving as seed sites for the galvanic displacement of $Ag^+$, the reaction (2):

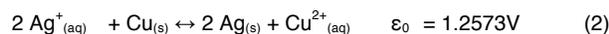

$$2\ Ag^+_{(aq)} + Cu_{(s)} \leftrightarrow 2\ Ag_{(s)} + Cu^{2+}_{(aq)} \quad \varepsilon_0 = 1.2573V \quad (2)$$

occurred gradually over 5-8 minutes. The choice of concentration of soluble cations (50 mM) caused reaction (2) to occur under diffusion-limited conditions, producing complex networks of silver fractals [34].

Upon completion, silver fractal networks were rinsed of any remaining reagents by dilution with a droplet (20 µL) of deionized water (18 MΩ). One half (20 µL) of the diluted solution was removed by pipette, and the process was repeated an additional five times. The solution remaining on the device was carefully removed with an absorbent wipe and then evaporated by placing the device on a hot plate at 70° C for 10 minutes. The entire rinsing process was completed with minimal disturbance to the network. As the networks dried, three-dimensional silver fractal structures flattened to a quasi-2D shape in which overlapping structures came into contact with each other, forming an interconnected network [33]. Electrical resistance of the silver fractal network ranged from 40-60 Ω.

To convert interconnections among Ag fractal structures into the MIM junction requisite for atomic switch operation, the networks were functionalized by exposure to sulfur gas. This sulfurization step was conducted by heating a crucible of sulfur (Sigma, 99.98%) to 140° C on a hot plate. The device was mounted onto a wire stand (10 cm) placed on the hot plate to ensure that the device remained close to room temperature during sulfurization. The sample, sulfur and wire stand were then covered with a large beaker for 10 minutes. The confined evaporated sulfur gas then deposited onto the cooler fractal network. The resulting thin film of sulfur reacted with the Ag fractal structures to form an insulating $Ag_2S$ surface layer while leaving the bulk internal structures as metallic Ag, creating an assemblage of Ag|$Ag_2S$|Ag MIM junctions. Electrical resistance of the across the ASN device was measured periodically to assess the degree of sulfurization. Devices with negligible resistance change from the initial state (40-60 Ω) were re-sulfurized to achieve appropriate values for memristive operation (0.5 - 1.5 MΩ).

Structural characterization was carried out using the optical (Nikon Eclipse TE2000-U) and scanning electron (FEI Nova600 NanoLab) microscopes. Electrical characterization employed current-voltage (*I-V*) spectroscopy using a precision source measure unit (National Instruments 4132) for resistance measurements, while timeseries IV data was collected using an analog voltage input/output module (National Instruments 6368) in conjunction with a current-to-voltage preamplifier (Stanford Research Systems SR570). Resistance measurements were conducted using a 2-electrode configuration with a 100 ms, 200 mV pulse. Timeseries data was collected at 10 kHz. Subsequent data analyses were carried out using MATLAB 2010b (MathWorks) and Origin 8.1 (OriginLab Corporation).

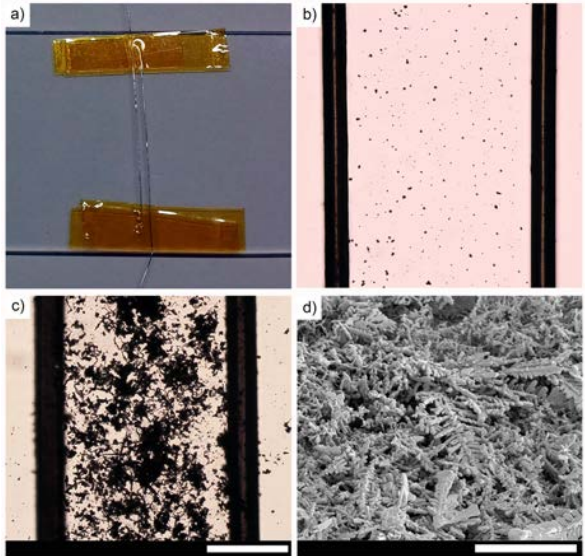

**Figure 3 - Device Fabrication – Illustrative representation of a device. (a)** Initial state of devices with both wires fastened to the glass slide. The possible sites of contact on the wires are insulated. **(b)** Copper spheres applied to the area between electrodes react with AgNO$_3$, **(c)** resulting in electroless deposition of complex silver nanofilament and fractal networks (scale bar = 1mm). **(d)** SEM image of silver fractals (scale bar = 30 $\mu$m).

**Results and Discussion.** In order to maximize both the number and proximity of MIM switching junctions, substrate coverage of dispersed copper microspheres was optimized in order to achieve a dense 2-D silver fractal network. Modelling and simulation of the resultant network topology was carried out using a tesselation method based on Voronoi diagrams in combination with a diffusion-limited model of electroless deposition. By partitioning a plane of $n$ points into convex polygons (called Voronoi Cells) such that each polygon contains exactly one nucleation site and every point in a given polygon is closer to its generating point than to any other, the resulting simulations provided direct insight into the design of fractal-based networks with respect to optimal surface coverage of microsphere seeds. As seen in Figure 3, metallic silver fractals generated by electroless deposition resulted in self-similar structures with fractal dimensions 1.72.

As-fabricated ASNs are initially collections of interconnected ohmic resistors. In common with many memristive devices, ASNs require an initial forming step to create a high conductivity 'ON' state through application of a sufficiently large bias voltage [30, 35]. Upon application of an appropriate activation bias voltage, subsequent sweeps produced an increased current output caused by a bias-induced phase transition from semiconducting acanthite ($\alpha$-Ag$_2$S) to conductive argentite ($\beta$-Ag$_2$S) within the MIM junctions alongside concurrent formation of conductive metal (Ag) filaments through reduction of soluble Ag cations at the cathode. This formed a distributed assembly of weakly memristive junctions in which electronic conduction occurred through the $\beta$-Ag$_s$S across a gap of decreased width due to the forming metallic filaments. Continued application of bias voltage induced filament growth across the gap, reducing the gap widths and increasing the measured current. The network then comprised resistive junctions with a growing number of completed memristive elements (atomic switches) and increasingly non-linear *I-V* behavior. The formation of a complete percolative path of memristive junctions from anode to cathode resulted in a sharp increase in network conductivity and defined the completion of this activation process.

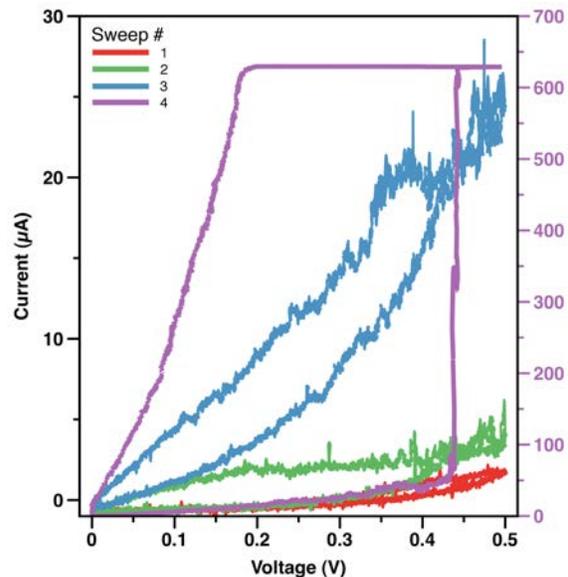

**Figure 4 - ASN device activation by applied bias voltage sweeps.** With each successive sweep (0-500 mV, 1 Hz) the resistance appeared to decrease quickly while voltage increased linearly and increase slowly while voltage decreased linearly. Continual increases in conductance (3.86 $\mu$S, 10.94 $\mu$S, 50.2 $\mu$S, 1.248 mS at peak voltage) are typical in initial forming steps for memristive devices as conducting channels through the ASNs lengthen and increase in number.

The entire forming step of an ASN device is shown in Figure 4 as a series of consecutive 0-500 mV sweeps, where an initially gradual increase in conductance was followed by an abrupt transition to the activated 'ON' state. Here, successive positive bias voltage sweeps of increasing amplitude were applied to initiate the forming step until measureable current began to flow through the ASN device. The minimum voltage required for activation ranged from 0.3 to 2 V. Following activation, ASN devices were stimulated with symmetric (±300 mV, 1-50 Hz) triangle wave voltage inputs. As Figure 5 shows, ASNs displayed repeatable switching between distinct resistance states, resulting in the pinched hysteresis *I-V* behavior typical of memristive systems. During consecutive sweeps, the threshold voltage for the low resistance state steadily decreased. Observed in individual switches, this behavior is due to a steady decrease in gap width where the theoretical minimum threshold is due to the oxidation and reduction of a single Ag atom which breaks and completes the filament [36]. In the network setting, reduction in threshold voltage is more likely due to a decrease in the ensemble average of gap widths in the ASN.

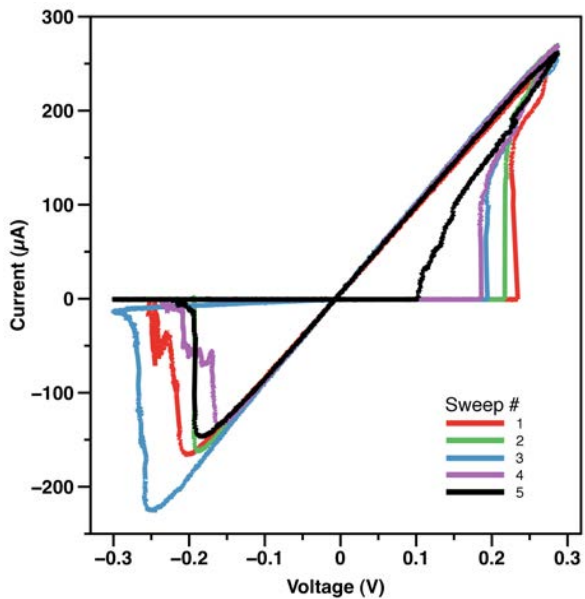

**Figure 5 – Robust hysteretic switching.** Five consecutive hysteresis curves show that each sweep is associated with a lower threshold voltage, due to decreasing insulator gap sizes. This data was produced directly after the network activation step. Voltage stimulus was a symmetric triangle wave (10 Hz, 300mV).

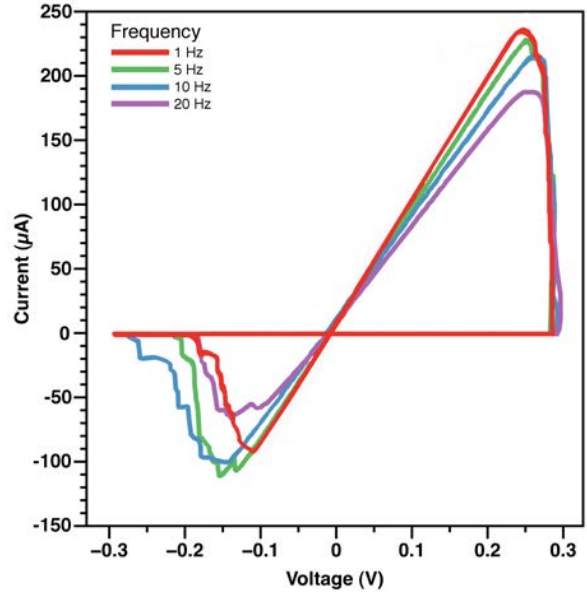

**Figure 6 – Frequency dependent hysteresis.** Representative average output of 10 sweeps with a given threshold voltage of 300 ±10 mV at 1 Hz, 5 Hz, 10 Hz and 20 Hz. Greater *I-V* hysteresis is associated with lower frequency sweeps. The amplitude of the current near the threshold value is typically greater for lower frequency sweeps. Conductances in increasing order of frequency were 1.01 mS, 0.94 mS, 0.72 mS and 0.57 mS.

A known fundamental property of memristive systems is that stimulation at increased frequency diminishes the degree of resistance change, resulting in more linear, ohmic *I-V* behavior [2, 6, 37]. As shown in Figure 6, this behavior was observed in ASN devices over an input bias frequency range of 1-20 Hz, where conductance decreased from 1.01 mS at 1 Hz to 0.57 mS at 20 Hz. Given constant input bias amplitude, a higher input frequency permits less net flux to pass through the junction during each half-sweep (i.e. 0 to +300 mV to 0). Charge carriers therefore have less time to migrate and are forced to reverse direction more often, causing them to instead fluctuate about their equilibrium point, resulting in larger gap widths and higher resistance. In contrast, slower sweeps allow enough cations to migrate not only to complete a conductive filament but to thicken it, which further increases conductance.

Additional evidence of frequency dependent behavior was observed alongside changes in the $R_{ON}/R_{OFF}$ ratio. Devices would occasionally fail to switch ON during application of AC bias voltage, remaining in the OFF state for the duration of a complete triangle wave voltage sweep. A minimum current output threshold ($I_t$ = 20 µA) was defined such that a complete sweep which failed to produce a current above $I_t$ and thus displayed no memristive properties, was deemed as lacking a switching event. Sweeps that displayed clear hysteresis and current output above $I_t$ were labeled as having a switching event. The device was then stimulated repeatedly for 60 second trials at 1, 5, 10, 20, or 50 Hz. The frequency for each trial was randomly selected in order to minimize the potential effects that a trial at a particular frequency would have on subsequent trials at different frequencies. As shown in Figure 7, the percentage of sweeps containing switching events was seen to decrease with increasing frequency of applied bias. This is qualitatively in agreement with the expected trend of *I-V* nonlinearity versus input frequency.

As compared to isolated atomic switches, ASNs require that gap widths be considered in the context of a network setting where filament creation and annihilation can result in complex interactions of many individual MIM junctions [30, 35]. At relatively low frequencies, repeatable switching between two primary distinguishable states shown in Figures 5 and 6 was dominated by the rapid transition to and from a low resistance state occurring immediately upon the completion or breakage of a completed percolative pathway across the network. It is therefore inferred from the data in Figure 7 that at higher frequencies, the amount of flux per cycle becomes progressively insufficient to allow for the completion of Ag filament formation. Thus the frequency dependent behavior is the result not only of the thickness of Ag filaments in the percolative pathway but also by whether a complete pathway is able to form.

**Conclusions and Outlook.** Using accessible materials and benchtop synthetic techniques we have prepared memristive atomic switch networks which display fundamental properties of memristive systems such as frequency-dependent, pinched hysteretic switching. The behavior of these devices is comparable to both theoretical predictions and previously reported properties of ASNs and memristive networks. The method of fabrication demonstrated here veers away from the top-down approach of complex lithographic techniques and instead utilizes a simplified, cost-effective method for the creation and study of memristive devices through bottom-up self-assembly. Memristors will undoubtedly find many uses as memory or logic elements in both conventional and neuro-inspired computing and new fabrication approaches such as this will aid in realizing their potential.

The simplicity of this benchtop approach to ASN fabrication also provides an accessible platform for further investigations of complex systems and an educational tool to introduce memristive systems. Because this method can produce easily reconfigurable networks, it provides an opportunity to investigate the properties of ASNs in various arrangements without requiring the development of unique designs or complex fabrication protocols. Recently it has been suggested that memristors or ASNs connected in parallel or series be studied [35]. A few proposed configurations for benchtop ASN devices which connect multiple networks in either series or parallel are provided in Figure 8. Redistribution of current in parallel configurations could allow for memristive elements with increased current output and power dissipation. Series configurations would produce larger functional networks, which may reveal advantages not seen in single ASNs including tunable threshold voltages and recurrent feedback [30]. This work represents a step forward in facilitation of such explorations by reducing the technical demands and monetary expense of current research in memristive systems, making it more readily accessible to a broader population of investigators.

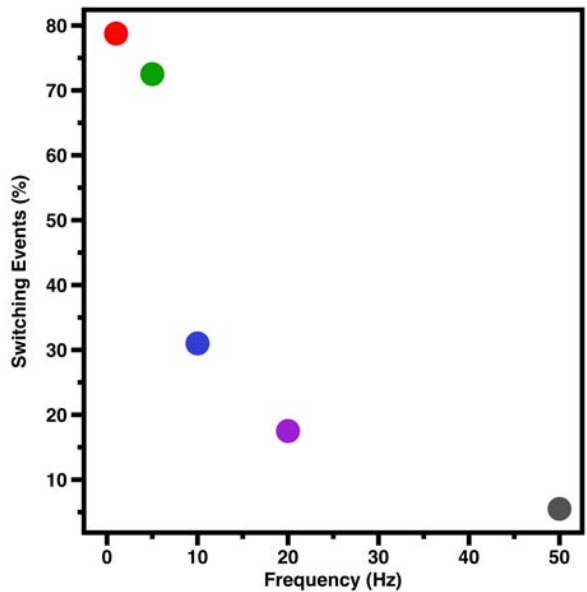

**Figure 7 - Ratio of switching events to total voltage sweeps as a function of frequencies.** Networks were stimulated with a symmetric triangle wave bias voltage (300 mV) at 1, 5, 10, 20, and 50 Hz. During a sweep, the device may display a switching event (pinched hysteresis) or it may remain in the high resistance state. The percentage of sweeps which contained a switching event was measured from an aggregate of over $10^3$ sweeps.

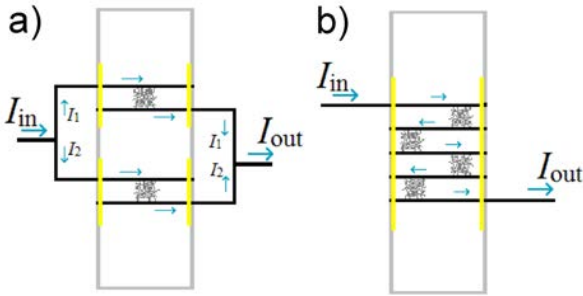

**Figure 8 - Schematic of parallel and series ASNs in benchtop devices**. The bottom-up fabrication approach allows for very flexible arrangements of networks. Additional networks may be connected in parallel (a) on an existing device by joining corresponding wires to the input and output nodes. Similarly, additional networks may be connected in series (b) by affixing wires to an existing device in the manner shown.

**Acknowledgements.** This work was partially supported by the Ministry of Education, Culture, Sports, Science, and Technology (MEXT) World Premier International (WPI) Research Center for Materials Nanoarchitectonics (MANA), HRL Laboratories, and the Defense Advanced Research Projects Agency (DARPA) - Physical Intelligence Program (BAA-09-63), US Department of Defense. The authors acknowledge use of the Nanoelectronics Research Facility (NRF) and Nano and Pico Characterization Laboratory (NPC) at the University of California, Los Angeles.

[*]Author to whom correspondence should be addressed: stieg@cnsi.ucla.edu

[†]Authors contributed equally to this work